\documentclass[sn-mathphys]{sn-jnl}

\usepackage{graphicx}%
\usepackage{amsmath,amssymb}
\usepackage[title]{appendix}%
\usepackage{xcolor}%
\usepackage{manyfoot}%
\usepackage{comment}
\usepackage{geometry}
\usepackage{paralist,tikz,pgfplots,comment}
\usetikzlibrary{matrix,calc,shapes}

\pgfplotsset{compat=newest}

\allowdisplaybreaks

\DeclareMathOperator{\diag}{diag}
\DeclareMathOperator{\tr}{tr}

\renewcommand{\vec}[1]{\boldsymbol{\mathrm{#1}}}
\newcommand*\ovec{\mathaccent"017E\relax}
\newcommand{\vvec}[1]{\overset{\tiny{{}_\leftarrow}}{#1}}

\begin{document}

\title[Article Title]{Gravitational anomaly detection using a satellite constellation: Analysis and simulation}


\author*[1]{\fnm{Viktor T.} \sur{Toth}}\email{vttoth@vttoth.com}


\affil*[1]{\orgaddress{\city{Ottawa}, \postcode{K1N~9H5}, \state{ON}, \country{Canada}}}


\abstract{We investigate the utility of a constellation of four satellites in heliocentric orbit, equipped with accurate means to measure intersatellite ranges, round-trip times and phases of signals coherently retransmitted between members of the constellation. Our goal is to reconstruct the measured trace of the gravitational gradient tensor as accurately as possible. Intersatellite ranges alone are not sufficient for its determination, as they do not account for any rotation of the satellite constellation, which introduces fictitious forces and accelerations. However, measuring signal round-trip time differences along clockwise and counterclockwise signal paths in a Sagnac-type measurement among the satellites supplies the necessary observables to estimate, and subtract, the effects of rotation. Utilizing, in addition, the approximate distance and direction from the Sun, it is possible to approach an accuracy of $10^{-24}~{\rm s}^{-2}$ for a constellation with typical intersatellite distances of 1,000~km in an orbit with a 1 astronomical unit semi-major axis. This is deemed sufficient to detect the presence of a galileonic modification of the solar gravitational field.}

\maketitle

\section{Introduction}

Recently, it has been proposed \citep{Yu2019} that an arrangement of four interplanetary satellites, orbiting the Sun in a tetrahedral configuration, may be used to measure the trace of the gradient tensor of the Newtonian gravitational field at sufficient accuracy to detect deviations from standard cosmology.

We investigated this proposal on its merits, by modeling the tetrahedral configuration in heliocentric orbit both analytically and numerically, through simulation. We find that the desired accuracy is, in fact, achievable but there are factors previously not considered that need to be taken into account. In particular, accounting for kinematic sources of acceleration (intrinsic rotation of the constellation or, to be more precise, the reference frame attached to the constellation in which relative accelerations are quantified) and accounting for higher-order tidal terms due to the gravitational field of the Sun are essential. We offer an analytic formulation of the desired observable, analyzing the sources and magnitudes of modeling errors. Our results are validated by a detailed simulation. Numerical limits are also considered as we conclude that a final, ``production quality'' model will necessarily require the use of extended precision arithmetic to achieve the desired precision. Nonetheless, the current simulation soundly confirms the feasibility of the approach as well as the analytical estimates of modeling errors.

We begin by reviewing the fundamentals of the modeled system in Section~\ref{sec:fund}. We introduce the concept of satellite-fixed reference frames, in which actual observations will be conducted, in Section~\ref{sec:frames}. In Section~\ref{sec:errors} we estimate both modeling and numerical errors and briefly discuss relativistic corrections. Our simulation results are presented in Section~\ref{sec:sim}. We summarize our conclusions in Section~\ref{sec:end}.

\section{Fundamentals}
\label{sec:fund}

The Newtonian gravitational field, characterized by its usual scalar potential $U$, obeys the gravitational Poisson equation \citep{Poisson1823}. The most general form of this equation for perfect fluids with density $\rho$ and isotropic pressure $p$ is given by\footnote{This form follows directly from the right-hand side of the alternate form of Einstein's field equation in the Newtonian approximation, specifically the term $T_{00}-\tfrac{1}{2}g_{00}T$ with $T_{\mu\nu}=\diag(c^2\rho,-p,-p,-p)$ for an isotropic perfect fluid.}
\begin{align}
\nabla^2 U=4\pi G(\rho+3c^{-2}p),
\end{align}
where $G$ is Newton's constant of gravitation and $c$ is the vacuum speed of light. Usually, the pressure term $p$ can be ignored unless the fluid is relativistic, resulting in the more commonly seen form $\nabla^2 U=4\pi G\rho$.

In the vacuum, $\rho=0$ hence $\nabla^2 U=0$. A perfect vacuum, of course, does not exist. For instance, in the interplanetary medium at $\sim 5$ particles per cubic centimeter in the vicinity of the Earth, assuming these to be protons yields
\begin{align}
\nabla^2 U\simeq 5.6\times 10^{-31}~{\rm s}^{-2}.
\end{align}
A relativistic example is provided by $p=-\rho$ dark energy in the standard cosmology; assuming a critical density $\rho_{\rm crit}\simeq 8.5\times 10^{-27}~{\rm kg}/{\rm m}^3$ and a dark energy density that is $\sim 70\%$ of the critical density, $\Omega_\Lambda\simeq 0.7$, we get
\begin{align}
\nabla^2 U_{\Lambda}\simeq -1.4\times 10^{-35}~{\rm s}^{-2}.
\end{align}
These numbers are very small, likely unobservable by available or foreseeable means. There are, however, proposed modifications of Newtonian gravitation in the form of ``galileon'' scalar fields \citep{Nicolis2009,Curtright2012}, which yield a small, $\propto r^{-1/2}$ deviation from the Newtonian gravitational acceleration. The corresponding effective potential, $U_g\propto \sqrt{r}$, does not obey the vacuum Poisson equation: $\nabla^2 U_g\ne 0$ in the vacuum. In \citep{Yu2019} the Laplacian is estimated at $\nabla^2 U_g\sim 10^{-24}~{\rm s}^{-2}$ at 1 AU (astronomical unit) from the Sun using cosmologically sensible parameterizations of the galileon theory. This quantity, albeit small, is presumed to be measurable by a tetrahedral configuration.

\subsection{The gravitational gradient tensor and its trace}

We characterize the gravitational field at the Newtonian level of approximation by its usual scalar potential $U$. The corresponding acceleration field that governs the motion of a field of non-interacting test particles is given by
\begin{align}
\vec{a}=\nabla U.
\end{align}
The gravitational gradient tensor is defined as
\begin{align}
\vec{T} = \nabla\otimes\nabla U,
\end{align}
where we use $\otimes$ for the outer product. The trace of the gradient tensor is
\begin{align}
\tr\vec{T}=\nabla^2 U,
\end{align}
and this is of course just the left-hand side of the gravitational Poisson-equation.

\subsection{Dynamics in the presence of the gradient tensor}

The gravitational gradient tensor can also be expressed as a function of the acceleration field:
\begin{align}
\vec{T}=\nabla\otimes\vec{a},
\end{align}
or equivalently, formally re-expressing this equation using infinitesimals,
\begin{align}
\vec{T}\cdot d\vec{r}=d\vec{a}.
\label{eq:Tdiff}
\end{align}
This can be extended to finite differences if $\vec{T}$ is expected to remain constant. Specifically, if we replace $d\vec{r}$ with $\Delta\vec{r}=\vec{r}_j-\vec{r}_i=\vec{r}_{ij}$, with $\vec{a}_{ij}$ as the corresponding accelerations, we have
\begin{align}
\vec{T}\cdot\vec{r}_{ij}=\vec{a}_{ij}.
\label{eq:Tconst}
\end{align}
If we have three distinct, non-parallel $\vec{r}_{ij}$, we can arrange them in a matrix $\vec{P}$, with the corresponding accelerations arranged in $\vec{A}$ and write
\begin{align}
\vec{T}\cdot\vec{P}=\vec{A},
\end{align}
which we can solve for $\vec{T}$:
\begin{align}
\vec{T}=\vec{A}\cdot\vec{P}^{-1}.
\end{align}
If $\vec{P}$ consists of three vectors $\vec{r}_{li}..\vec{r}_{lk}$, the inverse of $\vec{P}$ is given by
\begin{align}
\vec{P}^{-1}=\frac{1}{||\vec{P}||}\begin{bmatrix}{}[\vec{r}_{lj}\times\vec{r}_{lk}]^T\\{}[\vec{r}_{lk}\times\vec{r}_{li}]^T\\{}[\vec{r}_{li}\times\vec{r}_{lj}]^T\end{bmatrix},
\end{align}
with the determinant of $\vec{P}$ given by
\begin{align}
||\vec{P}||=\vec{r}_{li}\cdot[\vec{r}_{lj}\times\vec{r}_{lk}].
\end{align}
Thereafter, the trace of $\vec{T}$ is obtained as
\begin{align}
\tr\vec{T}=\sum_{ijk}\frac{\vec{a}_{li}\cdot[\vec{r}_{lj}\times\vec{r}_{lk}]}{\vec{r}_{li}\cdot[\vec{r}_{lj}\times\vec{r}_{lk}]},
\label{eq:trT}
\end{align}
where the summation is over all even permutations of $ijk$.

In addition to position vectors, we also require acceleration vectors to compute $\tr{\vec{T}}$. Given a time series of position vectors, the corresponding acceleration vectors can be constructed {\em in an inertial reference frame} the usual way:
\begin{align}
\vec{a}(t)\simeq\frac{\vec{r}(t-\Delta t)+\vec{r}(t+\Delta t)-2\vec{r}(t)}{\Delta t^2}.
\label{eq:approxa}
\end{align}

In a noninertial reference frame, this construction fails, as accelerations include the effects of fictitious forces.

\subsection{Second-order acceleration terms}

The method of computing $\tr{\vec{T}}$, in particular switching from the infinitesimal representation used in (\ref{eq:Tdiff}) to the finite differences captured in (\ref{eq:Tconst}), implies that we are accounting for acceleration field only to the first order with respect to its spatial gradient. In actuality, an acceleration field such as that of the Sun, includes higher-order components, which are not necessarily negligible.

To assess these contributions, let us consider the difference in acceleration for satellite $i$ vs. $j$ due to a central gravitating body such as the Sun:
\begin{align}
\vec{a}_{i} &{} = -\frac{GM}{|\vec{r}_i|^3}\vec{r}_i,\\
\vec{a}_{j} &{} = -\frac{GM}{|\vec{r}_j|^3}\vec{r}_j.
\end{align}
Using
\begin{align}
\vec{r}_{ij} &{} = \vec{r}_j - \vec{r}_i,
\end{align}
we can calculate
\begin{align}
|\vec{r}_{j}|^2 &{} = |\vec{r}_i|^2 + 2\vec{r}_i\cdot\vec{r}_{ij} + |\vec{r}_{ij}|^2 = |\vec{r}_i|^2\left[1 + \frac{2}{|\vec{r}_i|^2}\vec{r}_i\cdot\vec{r}_{ij}+\frac{|\vec{r}_{ij}|^2}{|\vec{r}_i|^2}\right],\\
|\vec{r}_{j}|^{-3} &{} \simeq |\vec{r}_i|^{-3}\left[1 - \frac{3}{|\vec{r}_i|^2}\vec{r}_i\cdot\vec{r}_{ij}+\frac{15}{2}\frac{(\vec{r}_i\cdot\vec{r}_{ij})^2}{|\vec{r}_i|^4}-\frac{3}{2}\frac{|\vec{r}_{ij}|^2}{|\vec{r}_i|^2}\right],\\
\vec{a}_{j} &{} = -\frac{GM}{|\vec{r}_i|^3}\left[1 - \frac{3}{|\vec{r}_i|^2}\vec{r}_i\cdot\vec{r}_{ij}+\frac{15}{2}\frac{(\vec{r}_i\cdot\vec{r}_{ij})^2}{|\vec{r}_i|^4}-\frac{3}{2}\frac{|\vec{r}_{ij}|^2}{|\vec{r}_i|^2}\right]\vec{r}_j,\nonumber\\
&{} = -\frac{GM}{|\vec{r}_i|^3}\left[\vec{r}_i+\vec{r}_{ij} - \frac{3}{|\vec{r}_i|^2}(\vec{r}_i\cdot\vec{r}_{ij})(\vec{r}_i+\vec{r}_{ij})+\frac{15}{2}\frac{(\vec{r}_i\cdot\vec{r}_{ij})^2}{|\vec{r}_i|^4}(\vec{r}_i+\vec{r}_{ij})-\frac{3}{2}\frac{|\vec{r}_{ij}|^2}{|\vec{r}_i|^2}(\vec{r}_i+\vec{r}_{ij})\right]\nonumber\\
&{} \simeq -\frac{GM}{|\vec{r}_i|^3}\left[\vec{r}_i+\vec{r}_{ij} - \frac{3}{|\vec{r}_i|^2}(\vec{r}_i\cdot\vec{r}_{ij})(\vec{r}_i+\vec{r}_{ij})+\frac{15}{2}\frac{(\vec{r}_i\cdot\vec{r}_{ij})^2}{|\vec{r}_i|^4}\vec{r}_i-\frac{3}{2}\frac{|\vec{r}_{ij}|^2}{|\vec{r}_i|^2}\vec{r}_i\right].
\end{align}
Using the identities,
\begin{align}
(\vec{r}_i\cdot\vec{r}_{ij})\vec{r}_i &{} = |\vec{r}_i|^2\vec{r}_{ij} - \vec{r}_i\times(\vec{r}_{ij}\times \vec{r}_i),\\
(\vec{r}_i\cdot\vec{r}_{ij})\vec{r}_{ij} &{} = |\vec{r}_{ij}|^2\vec{r}_i + \vec{r}_{ij}\times(\vec{r}_{ij}\times \vec{r}_i),
\end{align}
we can simplify this expression:
\begin{align}
\hskip -0.75em
\vec{a}_j &{} =-\frac{GM}{|\vec{r}_i|^3}\bigg\{\vec{r}_i+\vec{r}_{ij}-
\frac{3}{|\vec{r}_i|^2}\Big[|\vec{r}_i|^2\vec{r}_{ij} - \vec{r}_i\times(\vec{r}_{ij}\times \vec{r}_i) + |\vec{r}_{ij}|^2\vec{r}_i + \vec{r}_{ij}\times(\vec{r}_{ij}\times \vec{r}_i)\Big] \nonumber\\
&{}+\frac{15}{2}\frac{(\vec{r}_i\cdot\vec{r}_{ij})^2}{|\vec{r}_i|^4}\vec{r}_i - \frac{3}{2}\frac{|\vec{r}_{ij}|^2}{|\vec{r}_i|^2}\vec{r}_i
\bigg\}\\
&{} = -\frac{GM}{|\vec{r}_i|^3}\left[
\vec{r}_i
-2\vec{r}_{ij}
+\frac{3}{|\vec{r}_i|^2}\vec{r}_i\times(\vec{r}_{ij}\times\vec{r}_i)
-\frac{9}{2}\frac{|\vec{r}_{ij}|^2}{|\vec{r}_i|^2}\vec{r}_i
+\frac{15}{2}\frac{(\vec{r}_i\cdot\vec{r}_{ij})^2}{|\vec{r}_i|^4}\vec{r}_i
-\frac{3}{|\vec{r}_i|^2}\vec{r}_{ij}\times(\vec{r}_{ij}\times\vec{r}_i)\right],
\end{align}
allowing us to express the differential acceleration to second order as
\begin{align}
\vec{a}_{ij} &{} = -\frac{GM}{|\vec{r}_i|^3}\left[
-2\vec{r}_{ij}
+\frac{3}{|\vec{r}_i|^2}\vec{r}_i\times(\vec{r}_{ij}\times\vec{r}_i)
-\frac{9}{2}\frac{|\vec{r}_{ij}|^2}{|\vec{r}_i|^2}\vec{r}_i
+\frac{15}{2}\frac{(\vec{r}_i\cdot\vec{r}_{ij})^2}{|\vec{r}_i|^4}\vec{r}_i
-\frac{3}{|\vec{r}_i|^2}\vec{r}_{ij}\times(\vec{r}_{ij}\times\vec{r}_i)
\right]
\end{align}
The first two terms in square brackets in this expression are accounted for when we estimate acceleration using finite distances between satellites in accordance with (\ref{eq:Tconst}). The remaining terms are not.

Let us assume that we know the approximate Sun-to-constellation direction as $\vec{n}$ and the Sun-to-constellation distance as $r$. The difference, then, is
\begin{align}
\delta\vec{a}_{ij}\simeq-\frac{3GM}{r^4}\left[\frac{3}{2}|\vec{r}_{ij}|^2\vec{n}
-\frac{5}{2}(\vec{n}\cdot\vec{r}_{ij})^2\vec{n}
+\vec{r}_{ij}\times(\vec{r}_{ij}\times\vec{n})\right].
\label{eq:acorr}
\end{align}

Using this term to correct the estimated acceleration term when computing $\tr{\vec{T}}$ amounts to the recognition that the method of computing $\tr\vec{T}$ using (\ref{eq:trT}) is accurate only to first order, so rather than resulting in $\tr\vec{T}=0$, it results in a residual trace that corresponds to second-order contributions to the gravitational field. In the solar system, in a heliocentric orbit, these second-order contributions are expected to be dominated by solar gravity. This we can estimate at the required level accuracy using only the approximate direction to, and distance from, the Sun as seen from the constellation. As this removes any possible contribution due to the Sun, a nonzero $\tr\vec{T}$ is therefore indicative of a perturbation of the gravitational field due to some other origin.

\section{Reference frames and observables}
\label{sec:frames}

Our goal is to recover the gravitational gradient tensor using a constellation of four satellites forming a tetrahedron, using only observables that are available in-constellation, without relying on precision astrometry or radio-navigation.

To this effect, we assume that the system of satellites is configured with sufficiently accurate (e.g., laser) intersatellite ranging capabilities, which produce a time series of hextuplets of measurements: the lengths of the six tetrahedron edges. As we shall see, coherent retransmission and timing/phase measurements will also play an important role.

Apart from a chiral ambiguity, a tetrahedron can be reconstructed from the six range measurements. This can be seen easily if we consider that the four vertices of the tetrahedron can be represented by three coordinates each, for a total of twelve independent degrees of freedom; however, six of these degrees of freedom are absorbed into the choice of an arbitrary origin and orientation of a ``satellite-fixed'' coordinate reference frame. The remaining six degrees of freedom can be solved for from the six intersatellite ranges.

However, the resulting coordinate frame will not be inertial. The six intersatellite measurements do yield an accurate determination of the shape of the tetrahedron but not its motion or orientation relative to the ``fixed stars''. Nor is it guaranteed that the tetrahedron will move at uniform velocity or that its orientation remains constant in relation to the fixed stars. Any acceleration measurements made relative to the satellite-fixed reference frame, therefore, will include fictitious accelerations due to the pseudoforces present in a noninertial reference frame. It is necessary to identify and account for these fictitious accelerations before the tetrahedral configuration can be employed as a reliable instrument to measure the trace of the gravitational gradient tensor.

\subsection{Establishing a coordinate reference frame}

The discussion up to this point was generic: we made no assumption on the nature of the coordinate reference frame in which the relative satellite positions $\vec{r}_{ij}$ are expressed, other than noting that in general, this reference frame may be noninertial, and its rate of rotation must be accounted for.

For practical calculations it is important to offer an explicit implementation of a practical reference frame. Given a tetrahedral constellation of satellites, if the six intersatellite ranges can be determined from measurement, it is possible to establish precisely (up to a chiral ambiguity, as we shall see later) the shape of the tetrahedron that the four satellites form and affix a corresponding Cartesian coordinate system that is convenient to use.

\begin{figure}
\begin{center}
\begin{tikzpicture}
\draw[color=gray!25,line width=0.1pt] (-0.5cm,0.5cm) -- (4cm,2cm);
\draw[color=gray!25,line width=0.1pt] (-1cm,1cm) -- (3.5cm,2.5cm);
\draw[color=gray!25,line width=0.1pt] (-1.5cm,1.5cm) -- (3cm,3cm);
\draw[color=gray!25,line width=0.1pt] (-2cm,2cm) -- (2.5cm,3.5cm);

\draw[color=gray!25,line width=0.1pt] (0.75cm,0.25cm) -- (-1.75cm,2.75cm);
\draw[color=gray!25,line width=0.1pt] (1.5cm,0.5cm) -- (-1cm,3cm);
\draw[color=gray!25,line width=0.1pt] (2.25cm,0.75cm) -- (-0.25cm,3.25cm);
\draw[color=gray!25,line width=0.1pt] (3cm,1cm) -- (0.5cm,3.5cm);
\draw[color=gray!25,line width=0.1pt] (3.75cm,1.25cm) -- (1.25cm,3.75cm);

\draw[color=red!20,line width=0.5pt] (0,0) -- (-1cm,2cm);
\draw[color=red!20,line width=0.5pt] (0,0) -- (1.5cm,5cm);
\draw[color=red!20,line width=0.5pt] (3cm,1cm) -- (-1cm,2cm);
\draw[color=red!20,line width=0.5pt] (3cm,1cm) -- (1.5cm,5cm);
\draw[color=red!20,line width=0.5pt] (-1cm,2cm) -- (1.5cm,5cm);
\draw[->] (0,0) -> (4.5cm,1.5cm);
\draw[->] (0,0) -> (-2.5cm,2.5cm);
\draw[->] (0,0) -> (0cm,6cm);
\filldraw[color=red] (0,0) circle (0.1cm);
\filldraw[color=green] (3cm,1cm) circle (0.1cm);
\filldraw[color=blue] (-1cm,2cm) circle (0.1cm);
\filldraw[color=orange] (1.5cm,5cm) circle (0.1cm);
\node[anchor=west] at (4.5cm,1.5cm) {$x$};
\node[anchor=east] at (-2.5cm,2.5cm) {$y$};
\node[anchor=south] at (0cm,6cm) {$z$};
\node[anchor=north west] at (0,0) {$k$};
\node[anchor=north west] at (3cm,1cm) {$i$};
\node[anchor=south east] at (-1cm,2cm) {$j$};
\node[anchor=south west] at (1.5cm,5cm) {$l$};
\end{tikzpicture}
\end{center}
\caption{\label{fig:coords}The satellite-fixed coordinate system.}
\end{figure}
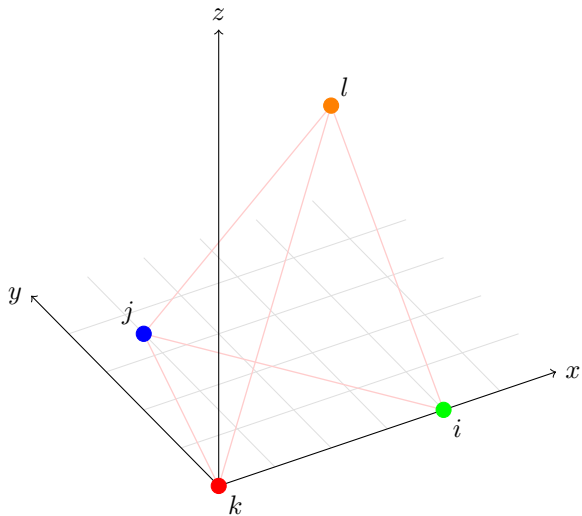

Our construction, shown in Figure~\ref{fig:coords}, is simple. We attach the origin of the coordinate system to one of the satellites, say, satellite $k$. We set the $x$ axis to correspond to the direction from satellite $k$ to satellite $i$. the $y$ axis, in turn, will be set to ensure that satellite $j$ remains in the $xy$-plane. The resulting basis vectors of our coordinate system are
\begin{align}
\vec{e}_x & {} = \frac{\vec{r}_{ki}}{|\vec{r}_{ki}|},\\
\vec{e}_z & {} = \frac{\vec{r}_{ki}\times\vec{r}_{kj}}{|\vec{r}_{ki}\times\vec{r}_{kj}|}\\
\vec{e}_y & {} = \vec{e}_z\times\vec{e}_x.
\end{align}

With this construction, expressing the intersatellite vectors in component form using only the six intersatellite range observables is reduced to a bit of straightforward algebra, leading to

\begin{align}
\vec{r}_k&{}=[0,0,0],\\
\vec{r}_i&{}=[r_{ki},0,0],\\
\vec{r}_j&{}=\left[\frac{r_{ki}^2 + r_{kj}^2 - r_{ij}^2}{2r_{ki}}, \sqrt{r_{kj}^2 - \left(\frac{r_{ki}^2 + r_{kj}^2 - r_{ij}^2}{2r_{ki}}\right)^2}, 0\right],\\
\vec{r}_l&{}=\begin{bmatrix}\dfrac{r_{ki}^2 + r_{kl}^2 - r_{il}^2}{2r_{ki}},\\~\\
 \dfrac{r_{kj}^2 + r_{kl}^2 - r_{jl}^2 - \dfrac{(r_{ki}^2 + r_{kl}^2 - r_{il}^2)(r_{ki}^2 + r_{kj}^2 - r_{ij}^2)}{2r_{ki}^2}}{2\sqrt{r_{kj}^2 - \left(\dfrac{r_{ki}^2 + r_{kj}^2 - r_{ij}^2}{2r_{ki}}\right)^2}},\\~\\
 \pm\sqrt{r_{kl}^2 - \left(\dfrac{r_{ki}^2 + r_{kl}^2 - r_{il}^2}{2r_{ki}}\right)^2 - \left(\dfrac{r_{kj}^2 + r_{kl}^2 - r_{jl}^2 - \dfrac{(r_{ki}^2 + r_{kl}^2 - r_{il}^2)(r_{ki}^2 + r_{kj}^2 - r_{ij}^2)}{2r_{ki}^2}}{2\sqrt{r_{kj}^2 - \left(\dfrac{r_{ki}^2 + r_{kj}^2 - r_{ij}^2}{2r_{ki}}\right)^2}}\right)^2}\end{bmatrix}.
\end{align}

We emphasize that our results do not in any way depend on the choice of this coordinate system. In particular, it does not matter which satellite we designate as satellite $k$. Other coordinate systems are also possible, including non-Cartesian coordinates, albeit in that case, care must be taken to ensure that expressions are evaluated properly; specifically, that cross-products are formed using the appropriate combination of the corresponding 3D metric tensor and Levi-Civita symbols (e.g., $\vec{u}\times\vec{v}=g^{\kappa\lambda}\epsilon_{\mu\nu\kappa}u^\mu u^\nu$.)

\subsection{Dealing with a rotating reference frame}

Given a tetrahedral constellation of four satellites, the six intersatellite ranges allow us to reconstruct the relative positions of the satellites. Picking a coordinate reference frame that is affixed to the satellite system, we can therefore construct position vectors $\vec{r}_i$ in component form. However, a satellite-fixed coordinate system is, in general, noninertial. Any accelerations measured in a noninertial frame will include fictitious accelerations, which must be removed. The relationship between an inertial and a noninertial system is characterized by the transformation
\begin{align}
\vec{r}'=\vec{R}(t)\cdot\vec{r}+\vec{x}(t),
\end{align}
where $\vec{R}(t)$ is a time-dependent, unitary rotation matrix and $\vec{x}(t)$ is a time-dependent vector. For differenced quantities, $\vec{x}(t)$ cancels:
\begin{align}
\vec{r}_{ij}'=\vec{r}_j'-\vec{r}_i'=\vec{R}(t)(\vec{r}_j-\vec{r}_i)=\vec{R}(t)\vec{r}_{ij}.
\end{align}
For inner products of differenced quantities, $\vec{R}$ also cancels:
\begin{align}
\vec{r}_{ij}'\cdot\vec{r}_{kl}'=(\vec{r}_{ij}\cdot\vec{R}^T)\cdot(\vec{R}\cdot\vec{r}_{kl})=\vec{r}_{ij}\cdot\vec{r}_{kl}.
\end{align}
Therefore, we can form inner products (and triple products) of such differenced vectors in any reference frame, including noninertial frames. However, the acceleration term in the numerator of $\tr\vec{T}$ in (\ref{eq:trT}) is a problem, because the acceleration term requires knowledge of $\vec{R}$:
\begin{align}
\vec{a}_{ij}'=\ddot{\vec{r}}_{ij}'=\vec{R}\cdot\ddot{\vec{r}}_{ij}+2\dot{\vec{R}}\cdot\dot{\vec{r}}_{ij}+\ddot{\vec{R}}\cdot\vec{r}_{ij}.
\end{align}
We know, however, that $\dot{\vec{R}}=\vec{R}\cdot\vec{\Omega}^T$, where the angular velocity tensor $\vec{\Omega}$, characterizing the rate of change of $\vec{R}$, is given by
\begin{align}
\vec{\Omega}=\begin{bmatrix}\vec{\omega}\times\vec{e}_1\\\vec{\omega}\times\vec{e}_2\\\vec{\omega}\times\vec{e}_3\end{bmatrix},
\end{align}
where $\vec{e}_i$ are the moving basis vectors of the coordinate frame. In a {\em Cartesian frame} it can be expressed using the components of the angular velocity $\vec{\omega}$:
\begin{align}
\vec{\Omega}=\begin{bmatrix}
0&-\omega_z&\phantom{-}\omega_y\\
\phantom{-}\omega_z&0&-\omega_x\\
-\omega_y&\phantom{-}\omega_x&0\end{bmatrix}.
\end{align}
We can thus write
\begin{align}
\vec{a}_{ij}'=\vec{R}\cdot(\ddot{\vec{r}}_{ij}+2\vec{\Omega}^T\cdot\dot{\vec{r}}_{ij}+(\vec{\Omega}^T)^2\cdot\vec{r}_{ij}+\dot{\vec{\Omega}}^T\cdot\vec{r}_{ij}),
\label{eq:arot}
\end{align}
and in the inner product between $\vec{a}_{ij}$ and $\vec{r}_{kl}$, $\vec{R}$ cancels:
\begin{align}
\vec{r}_{kl}'\cdot\vec{a}_{ij}'=(\vec{r}_{kl}\cdot\vec{R}^T)\cdot\vec{a}_{ij}'=\vec{r}_{kl}\cdot(\ddot{\vec{r}}_{ij}+2\vec{\Omega}^T\cdot\dot{\vec{r}}_{ij}+(\vec{\Omega}^T)^2\cdot\vec{r}_{ij}+\dot{\vec{\Omega}}\cdot\vec{r}_{ij}).
\label{eq:Krot}
\end{align}

We can always assume that the angular velocity is approximately constant, $\dot{\vec{\Omega}}=0$, by selecting a sufficiently small integration step in the time domain when numerically evaluating the system. With this assumption, we can extend the formalism above to compute the rotation of a vector $\vec{x}$ over a finite time interval $\Delta t$ using Rodrigues' formula \citep{Rodrigues1840}:
\begin{align}
\vec{x}'=(\vec{I}+(\sin\theta)\vec{K}+(1-\cos\theta)\vec{K}^2)\cdot\vec{x},
\end{align}
where $\vec{K}=|\vec{\omega}|^{-1}\vec{\Omega}$ and $\theta=|\vec{\omega}|\Delta t$. When $\Delta t$ is very small, this expression reduces to $\vec{x}'=(\vec{I}+\Delta t\vec{\Omega})\cdot\vec{x}$.

We solve for $\vec{\Omega}$ using a Sagnac-type \citep{Sagnac1,Sagnac2} observable of the time (or phase) difference between two signals sent around a triplet of satellites $kij$ in the clockwise vs. counterclockwise direction. In an nonrotating reference frame affixed to satellite $k$, the travel times of the three legs that this signal travels are given by
\begin{align}
ct_{ki}&{}=|\vec{r}_{ki}+t_{ki}\vec{v}_{ki}|,\label{eq:Sagnac1}\\
ct_{ij}&{}=|\vec{r}_{kj}+(t_{ki}+t_{ij})\vec{v}_{kj}-(\vec{r}_{ki}+t_{ki}\vec{v}_{ki})|,\\
ct_{jk}&{}=|\vec{r}_{kj}+(t_{ki}+t_{ij})\vec{v}_{kj}|,\label{eq:Sagnac3}
\end{align}
with $\vec{v}_{ij}=\dot{\vec{r}}_{ij}$. This system of equation is easily solvable for $t_{ki}$, $t_{ij}$ and $t_{jk}$:

\begin{align}
c^2t_{ki}^2&{}=r_{ki}^2+2\vec{r}_{ki}\cdot\vec{v}_{ki}t_{ki}+v_{ki}^2t_{ki}^2,\\
t_{ki}&{}=\frac{\vec{v}_{ki}\cdot\vec{r}_{ki}+\sqrt{(\vec{v}_{ki}\cdot\vec{r}_{ki})^2+(c^2-v_{ki}^2)r_{ki}^2}}{c^2-v_{ki}^2},\\
c^2t_{ij}^2&{}=[\vec{r}_{kj}-\vec{r}_{ki}+t_{ki}(\vec{v}_{kj}-\vec{v}_{ki})]^2+2[\vec{r}_{kj}-\vec{r}_{ki}+t_{ki}(\vec{v}_{kj}-\vec{v}_{ki})]\cdot\vec{v}_{kj}t_{ij}+v_{kj}^2t_{ij}^2,\nonumber\\
&{}=(\vec{r}_{ij}+t_{ki}\vec{v}_{ij})^2+2(\vec{r}_{ij}+t_{ki}\vec{v}_{ij})\cdot\vec{v}_{kj}t_{ij}+v_{kj}^2t_{ij}^2,\\
t_{ij}&{}=\frac{(\vec{r}_{ij}+t_{ki}\vec{v}_{ij})\cdot\vec{v}_{kj}+\sqrt{\{(\vec{r}_{ij}+t_{ki}\vec{v}_{ij})\cdot\vec{v}_{kj}\}^2+(c^2-v_{kj}^2)(\vec{r}_{ij}+t_{ki}\vec{v}_{ij})^2}}{c^2-v_{kj}^2},\\
t_{jk}&{}=\frac{\sqrt{(\vec{r}_{kj}+(t_{ki}+t_{ij})\vec{v}_{kj})^2}}{c}.
\end{align}

In a noninertial frame, however, using the fact that $t_{ij}$ are very small so we can treat $\vec{\Omega}$ representing the angular velocity as constant, we get, instead of (\ref{eq:Sagnac1})--(\ref{eq:Sagnac3}), the following equations:
\begin{align}
ct_{ki}'&{}=|\vec{r}_{ki}'+t_{ki}'(\vec{v}_{ki}'-\vec{\Omega}\cdot\vec{r}_{ki}')|,\label{eq:SagnacB1}\\
ct_{ij}'&{}=|\vec{r}_{kj}'+(t_{ki}'+t_{ij}')(\vec{v}_{kj}'-\vec{\Omega}\cdot\vec{r}_{kj}')-[\vec{r}_{ki}'+t_{ki}'(\vec{v}_{ki}'-\vec{\Omega}\cdot\vec{r}_{ki}')]|,\\
ct_{jk}'&{}=|\vec{r}_{kj}'+(t_{ki}'+t_{ij}')(\vec{v}_{kj}'-\vec{\Omega}\cdot\vec{r}_{kj}')|.\label{eq:SagnacB3}
\end{align}
The difference between the clockwise and counterclockwise timing measurement forms a Sagnac-type observable, which is given by
\begin{align}
\Delta t_{kij}=(t_{ki}+t_{ij}+t_{jk}) - (t_{kj}+t_{ji}+t_{ik}).
\end{align}
Observations in the noninertial satellite-fixed reference frame should reproduce this observable if the correct value of $\vec{\Omega}$ is used:
\begin{align}
\Delta t_{kij}=(t_{ki}'+t_{ij}'+t_{jk}') - (t_{kj}'+t_{ji}'+t_{ik}').\label{eq:DeltaSagnac}
\end{align}
At a given vertex of the tetrahedron, we have access to three such observables, corresponding to the three faces of the tetrahedron at that vertex. Using the resulting three equations, we can solve numerically for the three independent components of $\vec{\Omega}$, i.e., the angular velocity of the satellite-fixed reference frame. Using the solution for $\vec{\Omega}$, we can obtain $\vec{K}$. Using $\vec{K}$, we can correctly calculate accelerations in (\ref{eq:approxa}) by ``derotating'' the positions at $t-\Delta t$ and $t+\Delta t$ as appropriate:
\begin{align}
\vec{a}(t)=\frac{\vec{K}\cdot\vec{r}'(t-\Delta t)+\vec{K}^{-1}\cdot\vec{r}'(t+\Delta t)-2\vec{r}'(t)}{\Delta t^2}.
\end{align}
Consequently, we can compute the inner product between acceleration and position vectors that is needed to compute $\tr\vec{T}$.

\subsection{Accounting for acceleration terms}

The Sagnac-type observable developed in the preceding section, (\ref{eq:SagnacB1})--(\ref{eq:SagnacB3}), was modeled assuming that the corresponding relative satellite velocities can be treated as constant during a signal round-trip. As it turns out, this is indeed the case as the contribution of accelerations during the ${\cal O}(10^{-2}~{\rm s})$ signal round-trip times characteristic of a $\sim$1,000~km constellation are minuscule. Nonetheless, we also developed a formalism that incorporates accelerations. These can be accounted for by the expressions
\begin{align}
ct_{ki}&{}=|\vec{r}_{ki}+t_{ki}(\vec{v}_{ki}-\vec{\Omega}\cdot\vec{r}_{ki})+\tfrac{1}{2}t_{ki}^2(\vec{a}_{ki}+\vec{a}_k)|,\label{eq:SagnacC1}\\
ct_{ij}&{}=|\vec{r}_{kj}+(t_{ki}+t_{ij})(\vec{v}_{kj}-\vec{\Omega}\cdot\vec{r}_{kj})+\tfrac{1}{2}(t_{ki}+t_{ij})^2(\vec{a}_{kj}+\vec{a}_k)\nonumber\\
&{}-[\vec{r}_{ki}+t_{ki}(\vec{v}_{ki}-\vec{\Omega}\cdot\vec{r}_{ki})+\tfrac{1}{2}t_{ki}^2(\vec{a}_{ki}+\vec{a}_k)]|,\\
ct_{jk}&{}=|\vec{r}_{kj}+(t_{ki}+t_{ij})(\vec{v}_{kj}-\vec{\Omega}\cdot\vec{r}_{kj})+\tfrac{1}{2}(t_{ki}+t_{ij})^2(\vec{a}_{kj}+\vec{a}_k)-\tfrac{1}{2}(t_{ki}+t_{ij}+t_{jk})^2\vec{a}_k|,\label{eq:SagnacC3}
\end{align}
where, in addition to the relative accelerations $\vec{a}_{ij}$, we also included a term representing the absolute acceleration $\vec{a}_k$ of vertex $k$ representing the satellite that serves as the origin of the satellite-fixed reference frame. This acceleration is not known {\em a priori}. It turns out, however, that it is possible to solve for this acceleration if we have additional intrinsic observables on board, namely comparative signal round-trip times between pairs of satellites, namely the signal round-trip in the $kik$ and $iki$ direction between satellites $k$ and $i$. At any vertex $k$, three such measurements are available using the other three satellites; this yields a set of three equations in the three components of $\vec{a}_k$. Specifically, we have
\begin{align}
c\ovec{\tau}_{ki}&{}=|\vec{r}_{ki}+\ovec{\tau}_{ki}(\vec{v}_{ki}-\vec{\Omega}\cdot\vec{r}_{ki})+\tfrac{1}{2}\ovec{\tau}{\hskip 1pt}_{ki}^2(\vec{a}_{ki}+\vec{a})|,\\
c\ovec{\tau}_{ik}&{}=|\vec{r}_{ki}+t_{ki}(\vec{v}_{ki}-\vec{\Omega}\cdot\vec{r}_{ki})+\tfrac{1}{2}t_{ki}^2(\vec{a}_{ki}+\vec{a}) - \tfrac{1}{2}(\ovec{\tau}_{ki}+\ovec{\tau}_{ik})^2\vec{a}|,\\
c\vvec{\tau}_{ik}&{}=|\tfrac{1}{2}\vvec{\tau}{}_{ik}^2\vec{a}-\vec{r}_{ik}|,\\
c\vvec{\tau}_{ki}&{}=|\vec{r}_{ik}+(\vvec{\tau}_{ik}+\vvec{\tau}_{ki})(\vec{v}_{ki}-\vec{\Omega}\cdot\vec{r}_{ki})+\tfrac{1}{2}(\vvec{\tau}_{ik}+\vvec{\tau}_{ki})^2(\vec{a}_{ki}+\vec{a})-\tfrac{1}{2}\vvec{\tau}{}_{ik}^2\vec{a}|,\\
\Delta\tau_{ki}&{}=\ovec{\tau}_{ki}+\ovec{\tau}_{ik}-\vvec{\tau}_{ik}-\vvec{\tau}_{ki}.
\end{align}
The three differenced quantities $\Delta t_{kij}$, $\Delta t_{kjl}$ and $\Delta t_{kli}$ at vertex $k$, formed using (\ref{eq:DeltaSagnac}), along with the three quantities $\Delta\tau_{ki}$, $\Delta\tau_{kj}$, $\Delta\tau_{kl}$, form a set of six equations in the six unknown components of $\vec{\Omega}$ and $\vec{a}_k$.

Although no closed form solutions exist to this set of equations, they can be solved rapidly and efficiently by treating them as a pair of linked three-dimensional systems. Each of these systems can be solved using the Newton--Raphson method \citep{Raphson1697}, which in these cases yields rapid and robust convergence. The two solutions can be combined iteratively by a block Gauss--Seidel method \citep{Seidel1874}.

\section{Estimating errors}
\label{sec:errors}

A measurement of the type discussed here is subject to many different sources of error, that fall broadly into two categories: Systematic errors and random errors.

Systematic errors arise due to in-principle predictable but unmodeled characteristics of the system or instrument. Many sources of systematic errors exist in a realistic experiment. Here, we consider only one: the systematic error inherent due to the limitations of the mathematical model that we use to describe the four-satellite system and the gravitational environment, i.e., our modeling error.

Random errors may include stochastic errors due to uncontrolled sources of noise either external or intrinsic to the system, sampling noise, quantization noise, or rounding errors. Presently we only consider this last category, as the use of a finite-precision (standard double precision) number representation introduces a substantial error limiting the accuracy of our simulations.

\subsection{Modeling errors}

Our main observable is $\tr{\vec{T}}$, which has the dimensions of inverse time squared. This quantity is constructed using a vector triple product of relative accelerations and intersatellite ranges. While all measurement errors and uncertainties contribute, the largest contribution comes from the uncertainty due to the second-order acceleration term (\ref{eq:acorr}) that, as we have seen, is not fully accounted for when we reconstruct $\tr{\vec{T}}$ in the satellite-fixed reference frame from observable quantities only.

In particular, we can estimate the following first-order uncertainty in $\tr{\vec{T}}$, which would characterize this quantity if we omitted the second-order term (\ref{eq:acorr}) altogether:
\begin{align}
\delta_{[1]}\tr{\vec{T}}\simeq\frac{|\delta\vec{a}_{ij}|}{|\vec{r}_{ij}|}={\cal O}\left(GM\frac{d}{r^4}\right).
\end{align}
where we used $d$ to indicate the approximate scale of the satellite constellation. Using the gravitational parameter, $GM\simeq 1.3\times 10^{20}$~m$^3$/s$^2$ for the Sun and a $d\simeq 10^6$~m constellation at $r=1$~AU${}\simeq 1.5\times 10^{11}$~m from the Sun, we obtain
\begin{align}
\delta_{[1]}\tr{\vec{T}}\simeq\left(\frac{d}{1,000~{\rm km}}\right)\left(\frac{1~{\rm AU}}{a}\right)^4{\cal O}(3\times 10^{-19}~{\rm s}^{-2}),
\end{align}
where $a$ is the semi-major axis of the constellation's orbit.

Introducing the second-order tidal term (\ref{eq:acorr}) reduces the magnitude of this term by the factor $d/a$:
\begin{align}
\delta_{[2]}\tr{\vec{T}}\simeq\left(\frac{d}{1,000~{\rm km}}\right)^2\left(\frac{1~{\rm AU}}{a}\right)^5{\cal O}(2\times 10^{-24}~{\rm s}^{-2}),
\end{align}

We also encounter another source of error in the form of the Sagnac-type observable $\vec{\Omega}$ that we use to account for the noninertial pseudo-accelerations due to the rotation of the reference frame in which we model accelerations of the constellation. The equations (\ref{eq:SagnacB1})--(\ref{eq:SagnacB3}) do not account for the contribution of acceleration terms to $\vec{\Omega}$. The accelerations in question are relative (tidal) accelerations, $\vec{a}_{ij}={\cal O}(2GMd/a^3)={\cal O}(8\times 10^{-8}~{\rm m}/{\rm s}^2)$. Intersatellite signal travel times are $t_{ij}\simeq d/c={\cal O}(3\times 10^{-3}~{\rm s})$. Failure to include acceleration terms yields an uncertainty of $\delta(ct_{ij})=\tfrac{1}{2}t_{ij}^2\vec{a}_{ij}={\cal O}(4\times 10^{-13}~{\rm m})$ in the quantities $ct_{ij}$. This in turn implies an uncertainty in the angular velocity of order $\delta\omega=\delta(ct_{ij})/(t_{ij}|\vec{r}_{ij}|)={\cal O}(10^{-16}~{\rm s}^{-1})$.

According to (\ref{eq:arot}), we can estimate
\begin{align}
\delta|\vec{a}_{ij}| \simeq \delta|\omega\vec{v}_{ij} + \omega^2\vec{r}_{ij}|.
\end{align}
The two terms on the right-hand side are similar in magnitude. The constellation's intrinsic rotation is similar in magnitude to its orbital angular velocity, which implies, by Kepler's third law \citep{Kepler1619}, $\omega\propto a^{-3/2}$, thus
\begin{align}
\delta_{[\omega]}\tr{\vec{T}}\simeq\frac{|\delta\vec{a}_{ij}|}{|\vec{r}_{ij}|}\simeq\omega\delta\omega=
\simeq\left(\frac{d}{1,000~{\rm km}}\right)\left(\frac{1~{\rm AU}}{a}\right)^{4.5}{\cal O}(3\times 10^{-23}~{\rm s}^{-2}).
\end{align}

Finally, given $\tr\vec{T}\sim a_{ij}/r_{ij}$, considering a typical baseline of $10^6$~m at a 1~AU heliocentric orbit, to achieve a sensitivity of $\delta\tr\vec{T}\sim 10^{-21}~{\rm s}^2$, intersatellite ranges must be measured to a precision of ${\cal O}(10^{-5}~{\rm m})$ and accelerations at ${\cal O}(10^{-15}~{\rm m}/{\rm s}^2)$ or better. These precisions scale linearly with the desired sensitivity. This establishes the minimum required measurement accuracy.

\subsection{Numerical errors}
\label{sec:num}

In the preceding section, we computed the sensitivity limits of the proposed experiment. There are, however, also practical numerical limits that affect both modeling and data processing and will need to be addressed using appropriate numerical techniques. To wit, the standard double precision floating point number representation used on most computer hardware uses a 52-bit fractional mantissa, which offers just a little less than 16 decimal digits of effective precision. The more complex a calculation, the more its accuracy is reduced: multiple arithmetic operations represent a random walk away from the perfect answer, but subtraction of quantities of like magnitude can result in sudden significant drops in accuracy. As a general rule, for calculations as complex as the ones we employ for our current problem, the anticipated numerical accuracy of the result is around 14 decimal digits: $\delta^{\tt num}={\cal O}(10^{-14})$. Both of the previously calculated magnitudes of the accuracy of our results are affected by these limitations.

Differential (tidal) accelerations are calculated as differences of heliocentric accelerations. Consequently, the numerical error in our calculated result will be
\begin{align}
\delta^{\tt num}_{[\vec{a}]}\tr T\simeq \delta^{\tt num}\frac{GM}{da^2}=\left(\frac{1~{\rm AU}}{a}\right)^2\left(\frac{1,000~{\rm km}}{d}\right){\cal O}(6\times 10^{-23}~{\rm s}^{-2}).
\end{align}

Estimating $\omega$ runs into a more stringent numerical limitation. The Sagnac-type observable is calculated as a set of differences of the quantities (\ref{eq:SagnacB1})--(\ref{eq:SagnacB3}) and their counterclockwise counterparts. The quantities being subtracted have a magnitude of $3t_{ij}\simeq 3d/c={\cal O}(10^{-2}~{\rm s})$. The result of the subtraction, effectively the rotational displacement of a vertex during the signal round-trip time around the constellation, is $\Delta t\simeq 3t_{ij}\omega d/2c=3\omega d^2/2c^2={\cal O}(3\times 10^{-12}~{\rm s})$, which is ten orders of magnitude smaller. This implies the following limitation on the relative accuracy of $\omega$ when computed using double precision arithmetic:
\begin{align}
\frac{\delta\omega}{\omega}\simeq \delta^{\tt num}\frac{3t_{ij}}{\Delta t}=10^{-14}\frac{2c}{\omega d}={\cal O}(3\times 10^{-5}).
\end{align}
Consequently, we have
\begin{align}
\delta_{[\omega]}^{\tt num}\tr T\simeq \omega^2\frac{\delta\omega}{\omega}=10^{-14}\frac{2c\omega}{d} = \left(\frac{1~{\rm AU}}{a}\right)^{1.5}\left(\frac{1,000~{\rm km}}{d}\right) {\cal O}(1.2\times 10^{-18}~{\rm s}^{-2}).
\end{align}

These numerical limitations can be reliably addressed using extended precision arithmetic, at the cost of computational complexity and reduced computational speed.

\subsection{Comment on relativistic corrections}

The calculations that have been presented so far are all Newtonian. That is to say, these calculations implicitly assumed absolute time and a gravitational field adequately described by a Newtonian potential. Given a typical orbital velocity of $v\sim 30$~km/s, thus $(v/c)^2\sim 10^{-8}$, one may rightfully wonder if it is permissible to use the nonrelativistic approximation.

How may relativity theory affect our results? First, special relativistic effects concern coordinate transformations between moving reference frames. Second, general relativistic effects contribute to the gravitational acceleration. Let us address both these issues.

Concerning coordinate transformations, at first sight it may appear that these are indeed relevant. The orbital motion of the constellation may introduce centimeter-scale corrections to $d\sim 1000$~km intersatellite distances, with potentially significant impact on the results.

However, it is easy to see that this impact is readily absorbed in our formalism. Throughout, we implicitly assumed that intersatellite distances are expressed in a heliocentric inertial reference frame. This assumption was maintained even when we made use of these intersatellite distances in the satellite-fixed reference frame. This simply means that the instantaneous satellite-fixed reference frame used in our calculation is assumed to be at rest with respect to the heliocentric frame: that is to say, we moved the origin of the reference frame to one of the satellites, we rotated the reference frame, but we introduced no other motion (specifically removing any effects due to angular velocity by making use of the Sagnac-type observable).

In an actual experiment, this can be mimicked. Observations made on the satellites are fundamentally time observables, subject to relativistic time dilation. Knowledge of the satellite's motion around the Sun can be used to account for time dilation. This knowledge need not be precise: The (both special and general) relativistic effect being of ${\cal O}(v^2/c^2)$, an uncertainty $\delta v$ in $v$ introduces a relative error with the magnitude ${\cal O}(v\delta v/c^2)$. The magnitude of this error is further mitigated by the fact that in the end, our observables depend on not the absolute magnitude of intersatellite ranges but their differences. For the Sagnac-type observable this suppresses the contribution by a factor of $\sim 3\omega d/2c={\cal O}(10^{-9})$ as we have seen in Sec.~\ref{sec:num}. For the calculation of $\tr\vec{T}$, as we can see from (\ref{eq:trT}) we are scaling tidal acceleration per unit distance, $GM/a^3$, by $v\delta v/c^2$, yielding a contribution of $\sim 1.3\times 10^{-26}~{\rm s}^{-2}$ for $\delta v=1$~m/s, which is also negligible.

Therefore, we may conclude that as long as time observables are scaled by the appropriate $\gamma$-factor corresponding to the approximate heliocentric velocity of the constellation as well as the gravitational potential, even if it is not known to an accuracy better than a few ten m/s, additional relativistic corrections may be ignored. This justifies the nonrelativistic formalism used in our analysis. This is not to say that a more accurate analysis would not benefit from a fully relativistic formulation, but the results we present in our current study remain valid at our level of approximation.

\section{Simulation}
\label{sec:sim}

In the preceding sections, we built, from first principles, an analytical model of the tetrahedral configuration, estimated the model's accuracy, and explored methods by which the model's fidelity can be improved. Putting things into practice, we created a prototype implementation of the model in the form of a Web-based JavaScript application, complete with a simple user interface offering visualization by way of animation.

\subsection{Simulation code}

\begin{figure}
\begin{center}
\includegraphics{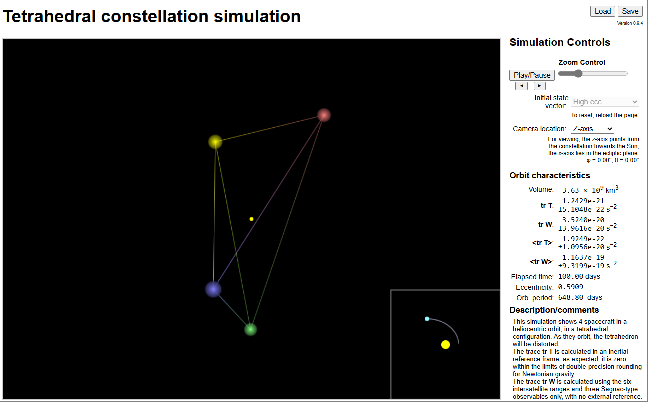}
\end{center}
\caption{\label{fig:ui}Screenshot of tetrahedral simulation of a $\sim 1$~AU eccentric orbit after 100 days.}
\end{figure}

To validate our analysis and results, we endeavored to build simulation code that implements the calculations presented so far. The simulation code was written in JavaScript, with a corresponding user interface in HTML5 and CSS. Source code for this implementation is available on GitHub\footnote{\url{https://github.com/vttoth/TETRA}}. An example screenshot is shown in Fig.~\ref{fig:ui}.

\begin{figure}

\pgfplotsset{width=2.8in,height=2.1in,compat=newest,
             every tick label/.append style={font=\footnotesize},
             error bars/error bar style={gray!75,no marks,opacity=0.2,ultra thin},
             error bars/error mark options={gray!05,opacity=0.0,rotate=90,ultra thin}}

\begin{tikzpicture}
\begin{axis}[
    error bars/.cd,
    xmin=0,
    xmax=650,
    ymin=-10,
    ymax=10,
    restrict y to domain*=-10:10,
]
\addplot[
    blue,
    no marks,
    error bars/.cd,
    y dir=both,
    y explicit,
] table[x expr=\thisrowno{0}, y expr=\thisrowno{2}*1e20, y error expr=((\thisrowno{3}>1e-19?1:(\thisrowno{3}<-1e-19?-1:\thisrowno{3}*1e20))), col sep=comma]{strLog.csv};
\addplot[
    blue,
    no marks,
] table[x expr=\thisrowno{0}, y expr=\thisrowno{2}*1e20, col sep=comma]{strLog.csv};
\node[blue] at (axis cs: 120,-3) {\footnotesize{$\times 10^{-20}~\rm{s}^{-2}$}};
\addplot[
  orange,
  dashed,
  no marks,
] table[x expr=\thisrowno{0}, y expr=\thisrowno{1}, col sep=comma] {strLog.csv};
\end{axis}
\end{tikzpicture}%
\begin{tikzpicture}
\begin{axis}[
    error bars/.cd,
    xmin=0,
    xmax=650,
    ymin=-10,
    ymax=10,
    restrict y to domain*=-10:10,
]
\addplot[
    blue,
    no marks,
    error bars/.cd,
    y dir=both,
    y explicit,
] table[x expr=\thisrowno{0}, y expr=\thisrowno{4}*1e19, y error expr=((\thisrowno{5}>1e-18?1:(\thisrowno{5}<-1e-18?-1:\thisrowno{5}*1e19))), col sep=comma]{strLog.csv};
\addplot[
    blue,
    no marks,
] table[x expr=\thisrowno{0}, y expr=\thisrowno{4}*1e19, col sep=comma]{strLog.csv};
\node[blue] at (axis cs: 150,-4) {\footnotesize{$\times 10^{-19}~\rm{s}^{-2}$}};
\node[orange] at (axis cs: 570,-4.5) {\footnotesize{$\times 10^{9}~\rm{km}^{3}$}};
\addplot[
  orange,
  dashed,
  no marks,
] table[x expr=\thisrowno{0}, y expr=\thisrowno{1}, col sep=comma] {strLog.csv};
\end{axis}
\end{tikzpicture}
\caption{\label{fig:pl}The value of $\tr\vec{T}$, as simulated (left) and as reconstructed from simulated observables (right) during one full orbit corresponding to the case shown in Fig.~\ref{fig:ui}. The variable (oriented) volume of the tetrahedron is also shown (dashed orange line), indicating that the uncertainty rises substantially when the tetrahedron becomes degenerate, its volume collapsing.}
\end{figure}
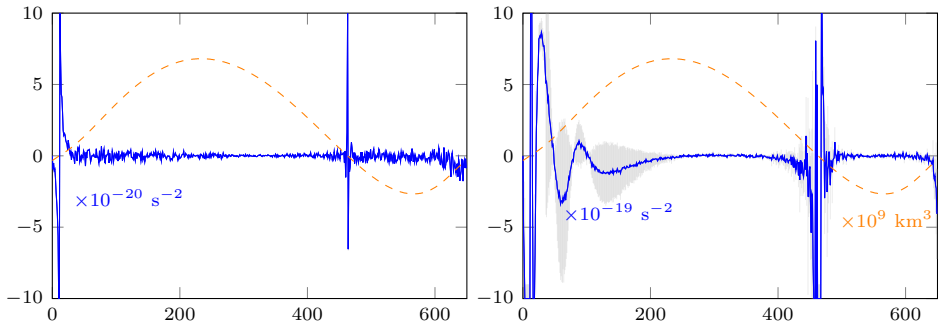

In this simulation, we use only the gravitational field of the Sun. Contributions from other solar system bodies, though present, should not alter our results as the vacuum Poisson equation ensures that $\nabla^2 U=0$; the gravitational field shapes the orbits of the satellites, but does not directly contribute to $\tr{\vec{T}}$ except through the approximations and omitted terms that we discussed. The contribution of solar system bodies other than the Sun to these terms is negligible.

The simulation code contains several preconfigured satellite constellations. Notable among them is an orbit with a relatively high eccentricity of $\epsilon=0.59$ and a perihelion of 0.6~AU. This orbit may be representative of a feasible experiment that might be carried out in the future. We also investigated other orbits, both near-circular and highly elliptical, including orbits with semi-major axes as small as $\sim 0.1$~AU and large as $\sim 30$~AU.

The JavaScript code performs three major functions, two of which are related to interacting with the user: managing user inputs and displaying results, and presenting an animated graphical representation of the four-satellite system.

At the core of the code, however, is the simulation itself that is used to model the four-satellite constellation at the maximum achievable accuracy using standard double precision floating point quantities. This ``physics package'' has two distinct but closely related aspects:
\begin{enumerate}
\item Modeling the physical behavior of the 4-satellite system;
\item Modeling the observation using only intersatellite measurements.
\end{enumerate}
The first of these two steps is performed in an inertial reference frame. To maximize the accuracy of the simulation, the origin of this reference frame is itself set to be moving with the satellite constellation, remaining within ${\cal O}(1000~{\rm km})$ of the satellites. This ensures that the integrated satellite positions remain accurate to $\sim 10^{-14}\times 1000~{\rm km}\sim 10^{-8}$~m, as opposed to the achievable accuracy using heliocentric coordinates in a double precision representation, $10^{-14}\times 1~{\rm AU}\sim 10^{-3}$~m.

To calculate orbits, we implemented a fourth order Runge-Kutta integrator, which we used to propagate satellite positions and velocities around the Sun. We also use a very tight integration step, e.g., 600~{\rm s} to integrate a $\sim$1~AU orbit, to ensure that the numerical integration remains as accurate as possible.

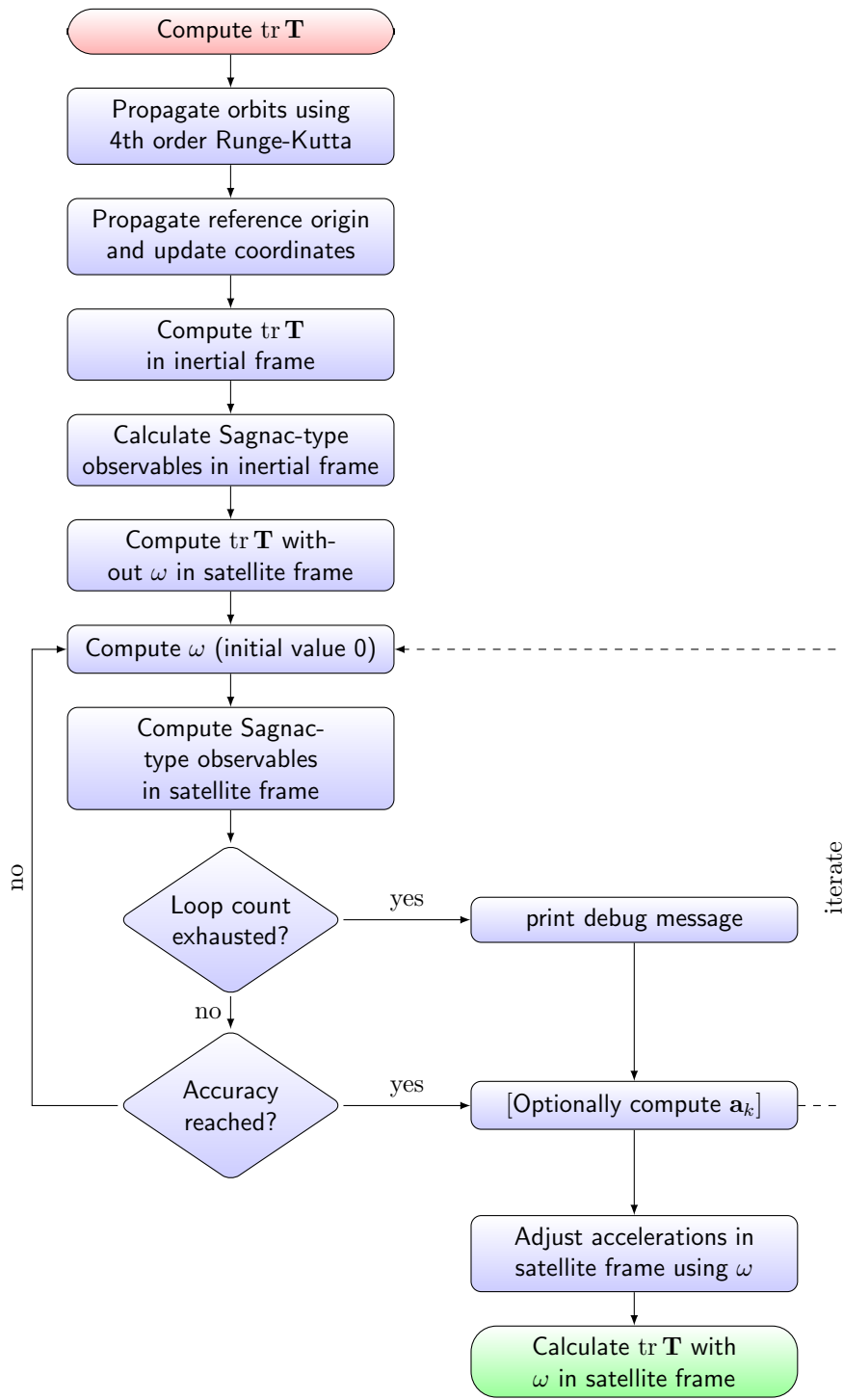
\begin{figure}
\begin{center}
\tikzset{
  treenode/.style = {shape=rectangle, rounded corners,
                     draw, anchor=center,
                     text width=12em, align=center,
                     top color=white, bottom color=blue!20,
                     inner sep=1ex, font=\sffamily},
  decision/.style = {treenode, text width=6em, aspect=1.5, diamond, inner sep=0pt},
  root/.style     = {treenode, font=\sffamily\normalsize, bottom color=red!30,rounded corners=10},
  env/.style      = {treenode, font=\ttfamily\normalsize},
  finish/.style   = {root, bottom color=green!40},
  dummy/.style    = {circle,draw}
}
\newcommand{\yes}{edge node [left] {yes}}
\newcommand{\no}{edge  node [left]  {no}}
\newcommand{\topyes}{edge node [above] {yes}}
\begin{tikzpicture}[-latex]
  \matrix (chart)
    [
      matrix of nodes,
      column sep      = 3em,
      row sep         = 3ex,
      column 1/.style = {nodes={treenode}},
      column 2/.style = {nodes={treenode}}
    ]
    {
      |[root]| Compute $\tr\vec{T}$                              & \\
      Propagate orbits using 4th order Runge-Kutta               & \\
      Propagate reference origin and update coordinates          & \\
      Compute $\tr\vec{T}$ in inertial frame                     & \\
      Calculate Sagnac-type observables in inertial frame        & \\
      Compute $\tr\vec{T}$ without $\omega$ in satellite frame   & \\
      Compute $\omega$ (initial value 0)                         & \\
      Compute Sagnac-type observables in satellite frame         & \\
      |[decision]| Loop count exhausted?                         & |[treenode]| print debug message \\
      |[decision]| Accuracy reached?                             & $[$Optionally compute $\vec{a}_k$$]$ \\
                                                                 & Adjust accelerations in satellite frame using $\omega$ \\
                                                                 & |[finish]| Calculate $\tr\vec{T}$ with $\omega$ in satellite frame \\
    };
  \draw
    \foreach \x/\y in {1/2, 2/3, 3/4, 4/5, 5/6, 6/7, 7/8, 8/9} {
    (chart-\x-1) edge (chart-\y-1) }
    \foreach \x/\y in {9/10, 10/11, 11/12} {
    (chart-\x-2) edge (chart-\y-2) }
    \foreach \x/\y in {9/10} {
      (chart-\x-1) \no (chart-\y-1) }
    \foreach \x in {9,10} {
      (chart-\x-1) \topyes (chart-\x-2) }
;
 \draw
   (chart-10-1) -- +(-2.75,0) |- (chart-7-1)
     node[near start,sloped,above] {no};
 \draw[dashed]
   (chart-10-2) -- +(+3,0) |- (chart-7-1)
     node[near start,sloped,above] {iterate};
   node[near start,above] {yes};
\end{tikzpicture}
\end{center}
\caption{\label{fig:flow}Flowchart illustrating the main computational loop in our simulation.}
\end{figure}

The core of the ``physics package'' follows tightly the logic depicted in the flowchart shown in Fig.~\ref{fig:flow}. A loop, which also controls updates to the graphical display, first propagates the satellite orbits and the coordinate origin and then computes the trace of the gravitational gradient tensor in the inertial frame. It also calculates the values of the Sagnac-type observables that characterize signal round trips around all four of the tetrahedron faces. Optionally, this computation takes into account intersatellite accelerations.

Next comes the critical part of the code, computation of the trace of the gravitational gradient tensor in the satellite-fixed reference frame. Initially, the rate at which this reference frame rotates is not known. However, the Sagnac-type observables can be calculated in the satellite-fixed frame. This is used to estimate the value of $\omega$, to ensure agreement with the same observables calculated in the inertial frame. The value of $\omega$ is obtained using an iterative numerical approximation. Optionally, the solution is extended by a second iteration that solves for the linear acceleration of the coordinate system origin.

Once the parameters, including the rate of rotation of the satellite-fixed frame, are known, the trace of the gravitational gradient tensor is recalculated.

These calculations are repeated for all four satellites. The trace of the gravitational gradient tensor is averaged across these four results and a standard deviation is calculated. Additionally, a continuously updated temporal average is also calculated, along with a corresponding standard deviation.

Ultimately, in addition to a visualization, the software shows a set of four values along with a corresponding set of four standard deviations: The instantaneous value of the trace of the gravitational gradient tensor as calculated in the inertial reference frame and as reconstructed using observables in the satellite-fixed frame; and a running temporal average of the same two values.

A few variables control the details of the simulation: whether or not to include second-order tidal corrections, use the Sagnac-type observable, include acceleration contributions to this observable, and calculate the linear acceleration of the origin of the satellite-fixed frame. Additionally, as a test case to validate the software, we implemented a Yukawa-type modification of the Newtonian gravitational potential, with a nonzero $\tr\vec{T}$.

\subsection{Results}

A snapshot of a representative case is shown in Fig.~\ref{fig:ui}, with characteristics of a full orbit in Fig.~\ref{fig:pl}. This image shows a snapshot of the simulation of the representative eccentric ${\cal O}(1~{\rm AU})$ orbit that corresponds to a realistic observing scenario. Simulation results that correspond to this scenario after 100 days of elapsed simulation time are summarized in Table~\ref{tab:results}. A companion table, Table~\ref{tab:averages}, shows the same scenario with results time-averaged over the first 100 days.

\begin{table}
\caption{\label{tab:results}A trial simulation of a constellation in an eccentric orbit, $\epsilon=0.59$ with a perihelion of 0.6~AU, with instantaneous values of $\tr\vec{T}$ after 100 days. Both ``true'' values (calculated in the inertial frame) and ``obs'' values (reconstructed from observables in the satellite-fixed frame) are shown.}
\begin{tabular}{l|c|c}
Description & $\tr\vec{T}_{\tt true} ({\rm s}^{-2})$ & $\tr\vec{T}_{\tt obs} ({\rm s}^{-2})$ \\
\hline\hline
Baseline run & $(-3.3369\pm 0.0051)\times 10^{-19} $ & $ (-1.0696 \pm 0.8423)\times 10^{-14} $\\
Baseline \& Sagnac & $(-3.3369\pm 0.0051)\times 10^{-19} $ & $(-0.3153 \pm 1.4796)\times 10^{-20} $\\
With 2nd tidal & $\phantom{-}(1.2429 \pm 0.5105)\times 10^{-21} $ & $ (-1.0696\pm 0.8423)\times 10^{-14} $\\
2nd tidal \& Sagnac & $\phantom{-}(1.2429\pm 0.5105)\times 10^{-21} $ & $\phantom{-}(3.5240\pm 3.9616)\times 10^{-20} $\\
2nd tidal, Sagnac+$\vec{a}$ & $\phantom{-}(1.2429\pm 0.5105)\times 10^{-21} $ & $\phantom{-}(2.6560 \pm 5.4441)\times 10^{-20} $\\
2nd tid., Sgn.+$\vec{a}$, lin.$\vec{a}$ & $\phantom{-}(1.2429\pm 0.5105)\times 10^{-21} $ & $\phantom{-}(3.0230 \pm 3.5768)\times 10^{-20} $ \\
\end{tabular}
\end{table}

\begin{table}
\caption{\label{tab:averages}The same trial simulation shown in Table~\ref{tab:results}, with time-averaged values of $\tr\vec{T}$ after 100 days in orbit.}
\begin{tabular}{l|c|c}
Description & $ \left<\tr\vec{T}_{\tt true}\right> ({\rm s}^{-2})$ & $ \left<\tr\vec{T}_{\tt obs}\right> ({\rm s}^{-2})$\\
\hline\hline
Baseline run & $ (-0.9968 \pm 1.8672) \times 10^{-18} $ & $ (-0.4078 \pm 1.3471)\times 10^{-12} $ \\
Baseline \& Sagnac & $ (-0.9968 \pm 1.8672) \times 10^{-21} $ & $(-1.1553 \pm 1.2070)\times 10^{-18} $ \\
With 2nd tidal & $ \phantom{-}(0.0192 \pm 1.0956) \times 10^{-20} $ & $ (-0.4078 \pm 1.4371) \times 10^{-12} $ \\
2nd tidal \& Sagnac & $\phantom{-}(0.0192 \pm 1.0956) \times 10^{-20} $ & $ \phantom{-}(1.1637\pm 9.3199)\times 10^{-19} $ \\
2nd tidal, Sagnac+$\vec{a}$ & $\phantom{-}(0.0192 \pm 1.0956) \times 10^{-20} $ & $\phantom{-}(1.1416 \pm 9.3817) \times 10^{-19} $ \\
2nd tid., Sgn.+$\vec{a}$, lin.$\vec{a}$ & $\phantom{-}(0.0192 \pm 1.0956) \times 10^{-20} $ & $\phantom{-}(1.0551 \pm 8.6592)\times 10^{-19} $ \\
\end{tabular}
\end{table}

In this table, we present the results of our simulation progressively turning on various simulation terms. First, a ``baseline'' case: The trace of the gravitational gradient tensor is computed in the inertial reference frame without second-order tidal terms, and in the satellite-fixed frame, without accounting for the rotation of the frame. We can see that the result in the inertial reference frame has reduced accuracy, as expected: Second-order tidal terms are important. As for the satellite-fixed frame, failure to account for that frame's rotation yields values that are effectively useless, up to ten orders of magnitude worse than the hoped-for accuracy of this experiment.

It is also interesting to note that although the value of the trace of the gravitational gradient tensor is nonzero, the standard deviation (calculated by iterating over the four satellites, using each as the origin of the satellite-fixed reference frame) is very tight. This suggests that the deviation from zero is not some random error: Rather, we are actually modeling the unaccounted-for contribution to the gravitational gradient tensor by those second-order tidal terms.

Turning on these terms dramatically changes the result in the inertial frame, taking $\tr\vec{T}$ much closer to zero. There is, however, little improvement in the value of the trace calculated in the satellite-fixed reference frame.

That changes when we introduce the Sagnac-type measurement from which the rotation rate of the satellite-fixed reference frame is estimated. As a result, the estimate of $\tr\vec{T}$ is now much closer to zero, approaching the numerical limitations of double precision arithmetic that we discussed previously.

Additional, marginal changes are achieved when we introduce acceleration-dependent terms in the model of the Sagnac-type observable, and when we account for the linear acceleration of the coordinate system origin. However, these changes amount to little more than numerical noise at the level of accuracy that can be achieved using the double precision representation.

We also investigated a variety of different orbits. These are shown in Table~\ref{tab:orbits}, with results after one full orbit; time-averaged results over that same full orbit are in Table~\ref{tab:orbavgs}. These cases are notable in that they include both circular and eccentric orbits, as well as orbits much larger than 1~AU. These latter cases, in particular, explicitly demonstrate how the anticipated accuracy with which $\tr\vec{T}\simeq 0$ can be verified increases as larger orbits are chosen. We recognize, of course, that orbits with very large heliocentric distances are not practical.

\begin{table}
\caption{\label{tab:orbits}Results for a set of additional orbits, identified by their orbital period $T$, eccentricity $\epsilon$ and average initial intersatellite range $\left<d\right>$, after one full orbit.}
\begin{tabular}{c|c|c|c|c}
$T$    & $\epsilon$ & $\left<d\right> $ & $ \tr\vec{T}_{\tt true} $ & $ \tr\vec{T}_{\tt obs} $\\
(days) & ~ & ($\times 10^3$~km) & $({\rm s}^{-2})$ & $({\rm s}^{-2})$ \\
\hline\hline
\phantom{00}373.35 & 0.0145 & 5.1\phantom{0} & $(-1.0346\pm 0.0323)\times 10^{-20}$ & $\phantom{-}(0.3964\pm 2.3749)\times 10^{-20}$ \\
\phantom{00}749.70 & 0.3808 & 3.7\phantom{0} & $(-2.3199\pm 0.5746)\times 10^{-21}$ & $(-1.4625\pm 2.3958)\times 10^{-19}$ \\
\phantom{00}648.80 & 0.5909 & 1.4\phantom{0} & $(-1.7436\pm 0.0271)\times 10^{-20}$ & $(-8.7312 \pm 5.1166)\times 10^{-20}$ \\
\phantom{00}648.80 & 0.5909 & 0.14           & $\phantom{-}(7.0714\pm 0.0154)\times 10^{-20}$ & $\phantom{-}(0.8805\pm 1.0846)\times 10^{-17}$ \\
\phantom{0}7989.33 & 0.2786 & 5.1\phantom{0} & $\phantom{-}(6.8411\pm 0.5730)\times 10^{-25}$ & $(-1.0703 \pm 6.7502)\times 10^{-22}$ \\
48377.88           & 0.1546 & 5.1\phantom{0} & $(-9.9706 \pm 0.0791)\times 10^{-25}$ & $(-1.9817\pm 1.9040)\times 10^{-23}$ \\
\end{tabular}
\end{table}

\begin{table}
\caption{\label{tab:orbavgs}As in Table~\ref{tab:orbits}, but time averaged over one full orbit.}
\begin{tabular}{c|c|c|c|c}
$T$    & $\epsilon$ & $\left<d\right> $ & $ \left<\tr\vec{T}_{\tt true}\right> $ & $ \left<\tr\vec{T}_{\tt obs}\right> $\\
(days) & ~ & ($\times 10^3$~km) & $({\rm s}^{-2})$ & $({\rm s}^{-2})$ \\
\hline\hline
\phantom{00}373.35 & 0.0145 & 5.1\phantom{0} & $(-0.1156\pm 3.8421)\times 10^{-21}$ & $(-0.2032\pm 1.4454)\times 10^{-20}$ \\
\phantom{00}749.70 & 0.3808 & 3.7\phantom{0} & $(-0.1237\pm 7.7005)\times 10^{-21}$ & $\phantom{-}(0.2773\pm 3.9310)\times 10^{-19}$ \\
\phantom{00}648.80 & 0.5909 & 1.4\phantom{0} & $(-0.0857\pm 6.0919)\times 10^{-21}$ & $(-0.0423 \pm 4.0115)\times 10^{-19}$ \\
\phantom{00}648.80 & 0.5909 & 0.14           & $(-0.0106 \pm 5.3436)\times 10^{-20}$ & $(-0.0034 \pm 1.4380)\times 10^{-17}$ \\
\phantom{0}7989.33 & 0.2786 & 5.1\phantom{0} & $(-2.3773 \pm 5.8368)\times 10^{-23}$ & $(-1.8261\pm 9.2527)\times 10^{-22}$ \\
48377.88           & 0.1546 & 5.1\phantom{0} & $(-0.0782 \pm 3.4742)\times 10^{-24}$ & $(-0.0386 \pm 1.0766)\times 10^{-22}$ \\
\end{tabular}
\end{table}

Though the actual results depend on many factors (e.g., the intersatellite ranges vary considerably during a full orbit, time averaging includes outliers, results with degraded accuracy when the tetrahedron volume briefly collapses) by and large these tables show that numerical results match our expected accuracy. Moreover, in all cases the primary limitation is due to the double precision floating point representation in this software implementation. A more accurate model using extended precision arithmetic is therefore called for to truly explore the limits of the utility of the tetrahedral configuration and find optimal orbits.

\subsection{Sanity check with Yukawa-gravity}

How do we know the validity of our calculations? In particular, how well does the reconstructed measurement follow the modeled value of $\tr\vec{T}$?

\begin{figure}[t]
\begin{center}
\includegraphics{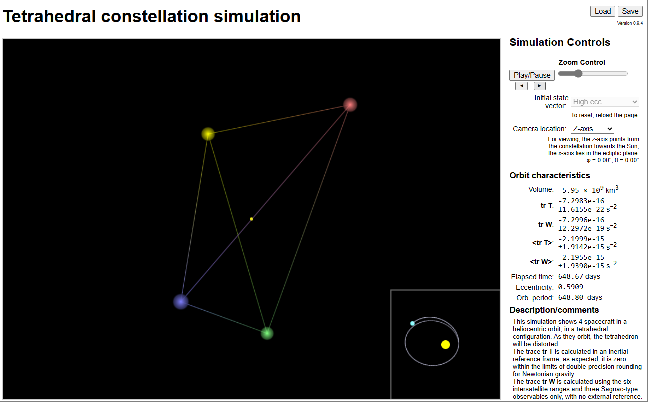}
\end{center}
\caption{\label{fig:yu}Screenshot of tetrahedral simulation with Yukawa-modified gravity. The large deviation chosen ($\gamma=0.2$, $\mu=(1~{\rm AU})^{-1}$) yields a visible precession after a full orbit. The orbital period shown in the user interface reflects the Keplerian value, not Yukawa gravity.}
\end{figure}

To answer these questions, we implemented a modification of Newtonian gravity by incorporating a Yukawa term:
\begin{align}
\vec{a}=-\frac{GM}{r^3}\vec{r}\qquad\Rightarrow\qquad
\vec{a}=-\frac{GM}{r^3}\left\{1+\gamma\left[1-(1+\mu r)e^{-\mu r}\right]\right\}\vec{r}.
\end{align}

An example run with Yukawa parameters $\gamma=0.2$, $\mu=(1~{\rm AU})^{-1}$, is shown in Figure~\ref{fig:yu}. These parameters are not intended to model any realistic modifications of gravity; they were used simply to validate the simulation code. As this screenshot illustrates, the reconstructed observable (marked $\tr\vec{W}$ in the user interface) closely follows the value modeled in the inertial reference frame. Temporal averaging is less well correlated but this has to do with outliers: over the course of more than a full orbit, the system went through stages of degeneracy when the tetrahedral volume was reduced to near zero, resulting in outliers that affected these averages.

\section{Conclusions}
\label{sec:end}

We investigated the use of a set of four satellites in near identical heliocentric orbits, forming a tetrahedral constellation, to be used as a means to measure deviations from Newtonian gravity. Such deviations are predicted by a variety of modified gravity theories, including galileon theories.

Specifically, the goal is to perform a local measurement of the trace of the gravitational gradient tensor, which is to say, the expression on the left-hand side of the gravitational Poisson equation $\nabla^2 U=4\pi G\rho$. In the vacuum, $\rho=0$, this trace should also be zero. The trace is measurable in principle by observing the dynamics of the system, since $\nabla U=\vec{a}$ forms an acceleration field, and $\nabla^2 U=\nabla\cdot\vec{a}$ at any point within the field.

We obtained a compact expression for the trace of the gravitational gradient tensor expressed using only the relative positions and relative accelerations of the satellites in the constellation. Inner products and triple products of relative position vectors are independent of the coordinate system in which these vectors are represented, therefore it is possible to compute them in a coordinate reference frame attached to the satellite constellation, without relying on any external references.

However, accelerations cannot be treated with such simplicity. Any reference frame attached to a moving constellation of satellites necessarily rotates. Relative accelerations, i.e., the second time derivatives of relative positions, can only be calculated in a rotating frame of reference if the frame's rate of rotation is known and accounted for. We realized that this rate of rotation can, in fact, be determined using measurements intrinsic to the constellation: Sagnac-type roundtrip timing measurements performed both in the clockwise and in the counterclockwise direction along any of the tetrahedron faces (i.e., any triplets of satellites). Together, three such measurements can be used to establish the rate of rotation of a coordinate frame in three-dimensional space.

The same equipment that is used for intersatellite range measurements (e.g., precision laser ranging) may be reused for the purpose of these Sagnac-type measurements. Thus, the rate of rotation of a satellite-fixed reference frame can be established without the need to rely on external references (be it astrometric measurements or the use of radio signals from ground stations) at accuracies that may not be practically achievable.

We also explored the limitations of the formalism to express the trace of the gravitational gradient tensor, and quickly recognized that second-order tidal effects, not directly accounted for by this formalism, are significant. The presence of these effects can substantially reduce the accuracy of the experiment. However, if we assume that such second-order tidal effects are due predominantly to the Sun, even approximate knowledge of the direction and distance towards the Sun, as expressed in the satellite-fixed reference frame, can be used to remove this contribution, substantially increasing the accuracy of the experiment.

We also looked at relativistic corrections and concluded that at the level of accuracy required for this experiment, moderately precise knowledge of the speed of the constellation relative to a heliocentric inertial reference frame is sufficient to correct on-board clock rates and obtain accurate results. We assumed that all on-board measurements include these corrections. Apart from these corrections, the use of a nonrelativistic model is fully justified.

Finally, we evaluated numerical limits. We found that the desired accuracy of measuring the trace of the gravitational gradient tensor up to ${\cal O}(10^{-24}~{\rm s}^{-2})$ is achievable. However, to reach this level of accuracy, models and simulations need to incorporate extended precision arithmetic, as double precision numbers are not sufficient.

To validate our effort, we implemented a simulation using the JavaScript programming language and a simple interactive user interface in HTML. Even though this model relied on double precision arithmetic only, thus it was limited in accuracy, it was used successfully to explore the parameter space. We found some satellite configurations that remained in a relatively stable tetrahedral configuration for one or several full orbits. Through these configurations, we were able to validate our predictions, demonstrate the fundamental feasibility of the experiment, and also validate our numerical expectations. We also validated the software by using it to model Yukawa-modified gravity, ascertaining that the system model and the modeled observable remain in good agreement.

The software model that we developed has been released as open source: \url{https://github.com/vttoth/TETRA}.

\backmatter

\bmhead{Acknowledgments}

VTT thanks Slava Turyshev for discussions and acknowledges the generous support of David H. Silver, Plamen Vasilev and other Patreon patrons.

\section*{Declarations}


\begin{itemize}
\item The authors declare that no funds, grants, or other support were received during the preparation of this manuscript.
\item The authors have no relevant financial or non-financial interests to disclose.
\item The authors consent to publication of this manuscript.
\item Code availability: The simulation code described in this manuscript is published at \url{https://github.com/vttoth/TETRA}.
\end{itemize}

\bibliography{refs}

\end{document}